\documentclass[aps,prd,showkeys,showpacs,amssymb,
amsfonts,epsf,preprintnumbers,nofootinbib,superscriptaddress]{revtex4}

\usepackage[dvips]{graphicx}
\usepackage{bm,latexsym,amsmath,amssymb,amsfonts,color}

\newcommand*{\tT}{{\tilde t}}

\newcommand*{\cT}{{\cal T}}
\newcommand\beq{\begin{eqnarray}}
\newcommand\eeq{\end{eqnarray}}

\newcommand{\nn}{\nonumber}

\newcommand{\cM}{\mathcal{M}}

\begin{document}

\title{Scalar field in the anisotropic universe}

\author{Hyeong-Chan Kim}
\email{hckim@cjnu.ac.kr}
\affiliation{Division of Liberal Arts, Chungju National University, Chungju 380-702,
  Korea}
\author{Masato Minamitsuji}
\email{minamituzi_at_sogang.ac.kr}
\affiliation{Center for Quantum Spacetime, Sogang University,
Shinsu-dong 1, Mapo-gu, Seoul, 121-742 South Korea}

\begin{abstract}

We discuss
the primordial spectrum
of a massless and minimally coupled scalar field,
produced during the initial
anisotropic epoch before the onset of inflation.
We consider
two models of the anisotropic cosmology,
the (planar) Kasner de Sitter solution (Bianchi I)
and the Taub-NUT de Sitter solution (Bianchi IX),
where
the 3-space geometry is initially anisotropic,
followed by the de Sitter phase
due to the presence of a positive cosmological constant.
We discuss the behavior of a quantized, massless
and minimally coupled scalar field
in the anisotropic stage.
This scalar field is not the inflaton
and hence does not contribute to the background dynamics.
We focus on the quantization procedure and evolution
in the pre-inflationary anisotropic background.
Also, in this paper
for simplicity the metric perturbations are not taken into account.
The initial condition is set by the requirement
that the scalar field is initially in an adiabatic state.
Usually, in a quantum harmonic oscillator system, 
an {\it adiabatic} process implies the
one where the potential changes slowly enough compared to its size,
and 
the time evolution can be obtained from the zero-th order
WKB approximation.
In our case, such a vacuum state exists only for
limited solutions of the anisotropic Universe,
whose spacetime structure is regular in the initial times.
In this paper,
we call our adiabatic vacuum state the {\it anisotropic vacuum}.
In the Kasner de Sitter model,
for one branch of planar solutions there is
an anisotropic vacuum unless $k_3\neq 0$,
where $k_3$ is the comoving momentum along the third direction,
while in the other branch there is no anisotropic vacuum state.
In the first branch,
for the moderate modes,
$k_3\sim k$, where $k$ is the total comoving momentum,
the scalar power spectrum has an oscillatory behavior
and its direction dependence is suppressed.
For the planar modes, $k_3\ll k$,
in contrast, the direction dependence becomes more important,
because of the amplification of the scalar amplitude
during this interval of the violation of WKB approximation
in the initial anisotropic stage.
The qualitative behaviors
in the Taub-NUT de Sitter models are very similar
to the case of the first branch of the planar Kasner de Sitter model.
\end{abstract}

\maketitle

\section{Introduction}

It is well-known that
our Universe is filled
with the thermal microwave radiation field,
so-called the cosmic microwave background (CMB).
CMB follows the almost idealized Planck distribution
of $2.73 K$.
The intensity of CMB is distributed almost isotropically,
but it contains small temperature fluctuations
of order $10^{-5}$ to the background temperature,
which contain much information about the history of the Universe.
Recent measurements by WMAP satellite
\cite{komatsu,wmap5,anomaly1} have shown
that the observed map of CMB anisotropy
is almost consistent
with the Gaussian and statistically isotropic
primordial fluctuations,
which is nothing but the most important predictions of
the inflation.

However, after the release of WMAP data,
several groups have reported
that there seem to be a few anomalies
in the CMB temperature map on large angular scales.
The most well-known fact is that there seems to
be suppression of the power of CMB fluctuations
on angular scales more than sixty degrees
\cite{anomaly1}.
There are other observational
facts that imply the effect which induces
the violation of the rotational invariance.
More precisely, probably there are
the planarity of lower multipole moments,
the alignment
between the quadrupole ($\ell=2$) and the octopole ($\ell=3$),
and the alignment
of them with the equinox and the ecliptic plane \cite{anomaly2}.
There are other observational facts implying the large-scale anisotropy, i.e.,
odd correlations of $\ell=4\sim 8$ multipoles with $\ell=2,3$ multipoles
\cite{copi},
a very large, possibly non-Gaussian cold spot in 10 degree scale
\cite{cruz},
asymmetry of angular map measured in north and south hemispheres
\cite{eriksen},
even though some authors claim that there is no significant evidence for primordial isotropy breaking in five-year WMAP data~\cite{Picon}.

Indeed, to explain the origin of the anomalies, various
solutions have been suggested,
introducing a non-trivial topology \cite{topology},
a locally anisotropy
based on the Bianchi type VII${}_h$ universe
to explain the 	
quadrupole/octopole planarity and alignment \cite{jaffe},
non-linear inhomogeneities \cite{moffat}
and assuming an elliptic Universe
to explain the suppression of
the quadrupole CMB power \cite{eli}.
More recently, in particular,
models which introduce an explicit source
to break
the spatial isotropy,
either during inflation or
in the late time Universe,
have been proposed,
e.g.,
by the dynamics of an
anisotropic energy-momentum component during inflation \cite{ack,anomaly,ys,wks},
by the large scale magnetic field \cite{vector},
by the anisotropic cosmological constant \cite{cc}
or dark energy \cite{de}.

In this paper, we will revisit
the possibility that such large scale anomalies
may be the relics of the physics of
preinflationary anisotropic Universe.
Wald's no-hair theorem ensures
that in the presence of a positive cosmological constant
an initially anisotropic Universe exponentially
approaches the de Sitter spacetime at the later time
under the strong or dominant energy condition \cite{wald}.
It implies that it is plausible that
the Universe is highly anisotropic {at the beginning}.
The cosmological perturbation theory in the pre-inflationary
Kasner phase {was} formulated in Ref. \cite{tpu,gkp}.
They showed
that in general in an expanding (planar) Kasner phase
one of two polarizations of gravitational waves
\footnote{Here, the {\it tensor} or {\it scalar} modes
are defined in the isotropic limit,
where perturbations are decomposed into modes on the
maximally symmetric 3-space.}
is coupled with the scalar mode,
but the other gravitational mode is decoupled.
Of course, the tensor-scalar coupling vanishes in the isotropic limit.
The latter mode can be amplified significantly before the onset of inflation.
The instability of the gravitational wave mode may
be deeply connected to the unstable mode
found by the Belinskii-Khalatnikov-Lifshitz (BKL) analysis \cite{bkl}
(see e.g., \cite{hnx} for review).
In a contracting Kasner universe such an instability
can be identified with the large scale anisotropic mode.

The problem is how to set the initial conditions.
In the standard inflation,
the initial condition is set
inside the Hubble horizon,
where the effects of the cosmic expansion can be ignored
and the adiabatic condition is satisfied.
Therefore, to compare with the prediction from the standard inflation,
it is natural to apply
similar arguments to the case of the anisotropic Universe.
Usually, in a quantum harmonic oscillator system,
an {\it adiabatic} process implies the
one where the potential changes slowly enough compared to its size,
and 
the time evolution can be obtained from the zero-th order
WKB approximation.
In this paper, we follow this definition for the term {\it adiabatic}. 
In the standard inflationary models,
an adiabatic vacuum is also defined in the same way. 
Note that an {\it isentropic} process implies 
$(adiabatic) + (reversible)$ process
and therefore is slightly different from our {\it adiabatic} one.  
We call our adiabatic vacuum state {\it an anisotropic vacuum}.
In an expanding Kasner solution,
there are two branches of solutions with the planar symmetry.
Then, the answer to the question on the initial adiabaticity
depends on the choice of the branch.
In one branch,
initially
the expansion rate along the planar directions vanishes
while that along the third axis is finite.
In this branch,
the initial spacetime structure can be seen as
(a patch of) the Minkowski spacetime,
and thus the corresponding anisotropic vacuum
has the physical meaning.
The tensor-scalar coupling mentioned above vanishes, and
therefore
the initial dynamics reduces to
the three-independent harmonic oscillators.
However, in the other branch
the tensor-scalar coupling diverges
{in going back to the initial time}.
In this branch, there is no adiabatic state.
Thus, we mainly focus on the first branch.
For a given set of initial conditions,
the power spectrum was investigated, resorting
to the numerical ways in Ref. \cite{tpu,gkp}.
One of the important motivations of this paper is
to aim more analytic understanding of the primordial power spectrum and
in particular its direction-dependence.
As the first step,
we will consider a free scalar theory,
which is the counterpart of the inflaton fluctuation.
The other important purpose is
to investigate the case of the other Bianchi model
and to see how generic the prediction of the Bianchi I model is.
In particular,
we will consider the analytic solution of the Bianchi IX Universe,
i.e., the Taub-NUT de Sitter solution.

The paper is constructed as follows:
In Sec. II, the background solutions in the Bianchi I and Bianchi IX models
are introduced,
say Kasner de Sitter solutions and Taub-NUT de Sitter solutions, respectively.
In Sec. III,
we investigate the behavior of a massless and minimally coupled scalar field
in the background of Kasner de Sitter solution with the planar symmetries.
We discuss the existence of a well-defined anisotropic vacuum for each mode.
in terms of the validity of the WKB approximation in the very early Universe
and determine the initial mode functions.
For the modes for which the anisotropic vacuum can be well-defined,
we derive the final power spectrum and investigate its direction dependence.
In Sec. IV, we repeat the similar discussions in the case of the
Taub-NUT de Sitter solutions.
In Sec. V, we close the article after giving a brief summary and discussion.

\section{Set-up}

We are interested in the
cosmological solutions of the Einstein gravity with a positive
cosmological constant in the absence of
the matter field with the action
\beq
S=M_{\rm Pl}^2\int d^4x \sqrt{-g}
\Big(\frac{1}{2}R-\Lambda\Big)\,,
\eeq
where $g_{\mu\nu}$ is the spacetime metric,
$\Lambda$ is a (positive) cosmological constant
and $M_{\rm Pl}$ is the Planck mass.

\subsection{Kasner-de Sitter spacetime}
Assuming the spatial part to be flat and homogeneous,
the solution to the Einstein equation is given by
\beq
ds^2=-d\tau^2
+e^{2\alpha(\tau)}
\Big[
 e^{2(\beta_++\sqrt{3}\beta_-)}(dx^1)^2
+ e^{2(\beta_+-\sqrt{3}\beta_-)}(dx^2)^2
+e^{-4\beta_+}(dx^3)^2
\Big],
\eeq
where
\beq
&&\alpha=\ln\Big(a_0\sinh^{1/3}\Big(3H\tau\Big)\Big),
\nonumber\\
&&\beta_+=\frac{m}{3} \ln\Big(\tanh\Big(\frac{3H\tau}{2}\Big)\Big),
\nonumber\\
&&\beta_-=\frac{1}{3}
\sqrt{1-m^2} \ln\Big(\tanh\Big(\frac{3H \tau}{2}\Big)\Big),
\eeq
{and}
$|m|\leq 1$ and $H:=\sqrt{\Lambda/3}$.
The solution initially behaves as a Kasner spacetime,
followed by a de Sitter phase,
as discussed in \cite{tpu,gkp}.
At the initial time
$\tau=0$, there is the Big Bang singularity.
In the later discussion, we will set $a_0=1$,
which can always be done by an appropriate rescaling of
the spatial coordinates $x^i$ ($i=1,2,3$).

In the later discussions, we will focus on the case
with a planar symmetry $m=+1$ or $m=-1$.
In the branch of $m=-1$, in the early time limit $\tau\to 0$
the scale factor in the planar direction
$e^{\alpha+\beta_+}$ approaches a constant, while
that in the $x^3$ direction $e^{\alpha-2\beta_+}$ linearly depends on $\tau$.
Thus, in this branch, there is one dynamical direction in the early time.
In the branch of $m=+1$, in the early time limit,
$e^{\alpha+\beta_+}\sim \tau^{2/3}$
and
$e^{\alpha-2\beta_+}\sim \tau^{-1/3}$,
and thus there are initially expanding planar directions
and contracting $x^3$ direction.
Thus, both the directions are dynamical.

\subsection{Taub-NUT de Sitter spacetime}

A Bianchi IX universe has three Killing vectors
$\xi_i$ satisfying $[\xi_i,\xi_j]=\epsilon_{kij}\xi_k$.
The corresponding one form $\chi^i=
(-\sin x^3 dx^1 +\sin x^1 \cos x^3 dx^2,
\cos x^3 dx^1+\sin x^1\sin x^3 dx^2,\cos x^1 dx^2+dx^3)$ constitutes the
{spatial} part of the metric.
Therefore, the general spacetime metric of a Bianchi IX universe is given by
\beq \label{met1}
ds^2=-n(t)^2 dt^2+a(t)^2\big(\chi^1\big)^2
+b(t)^2\big(\chi^2\big)^2
+c(t)^2\big(\chi^3\big)^2.
\eeq
For the case of {a} {\it planar} symmetry
in the sense of Bianchi I model, $a(t)=b(t)$,
the metric Eq. (\ref{met1}) becomes
\beq
ds^2=-n(t)^2 dt^2
+a(t)^2\Big(\big(dx^1\big)^2 +\sin^2 x^1 \big(dx^2\big)^2 \Big)
+c(t)^2\Big(dx^3+\cos x^1 dx^2\Big)^2\,.\label{met2}
\eeq
As a background solution, the 
Taub-NUT de Sitter solution is
considered
\beq
n(t)^2=\frac{t^2+\ell^2}{4\Delta(t)},\quad
a(t)^2=\frac{t^2+\ell^2}{4},\quad
c(t)^2=\frac{\ell^2 \Delta (t)}{t^2+\ell^2}\,,
\eeq
where
\beq
\Delta(t):=-t^2+2{\cal M}t +\ell^2
-\frac{\Lambda}{4}\Big(\ell^4-2\ell^2 t^2-\frac{t^4}{3}\Big)\,.
\label{def_del}
\eeq
In the case of $\Lambda=0$, the solution reduces to the Taub-NUT solution.
In the late time limit $t\to \infty$, the spacetime approaches de Sitter (dS)
with the Hubble expansion rate $H=(\Lambda/3)^{1/2}$.
As for the Taub-NUT space,
in the region $\Delta(t)<0$ where $t$ is spacelike,
the geometry is geodesic incomplete.
In this work,
we focus on the region where $\Delta(t)\geq 0$.
Requiring that there is no remaining anisotropy
in the late time Universe,
we choose $\ell^2\Lambda=3$
and obtain
\beq
\Delta(t)
=\frac{1}{4\ell^2}
\Big(
t^4+2\ell^2 t^2+8{\cal M}\ell^2 t+\ell^4
\Big)
\,.
\eeq
The function $\Delta(t)$ can be expressed
as
\beq
\Delta(t)=\frac{1}{4\ell^2}\big(t-t_0\big)
 \big(t^3+t_0t^2+(2\ell^2+t_0^2)t+(t_0^3+2\ell^2t_0+8{\cal M}\ell^2)),
\eeq
where $t_0$ satisfies
\beq
t_0^4+2\ell^2 t_0^2+8{\cal M}\ell^2 t_0+\ell^4=0.\label{ht}
\eeq
For $|{\cal M|}<2\ell/(3\sqrt{3})$,
there is no real root for Eq. (\ref{ht})
and for $|{\cal M}|>2\ell/(3\sqrt{3})$
there are two real roots
(For ${\cal M}>2\ell/(2\sqrt{3})$, $t_0$
is negative and for ${\cal M}<-2\ell/(3\sqrt{3})$,
$t_0$ is positive).
For $|{\cal M}|< 2\ell/(3\sqrt{3})$,
$\Delta$ is always positive,
and the spacetime geometry approaches de Sitter
in the past and future $t\to\pm\infty$,
while it is anisotropic during the intermediate time.
In the later discussions,
we will focus on the case of ${\cal M}\geq 2\ell/(3\sqrt{3})$, where
$\Delta(T)$ vanishes at two points,
represented by $t=t_{0,+},t_{0,-}$, where $t_{0,+}>t_{0,-}$
($t_{0,+}=t_{0,-}$ for ${\cal M}=\pm 2\ell/3\sqrt{3}$).
We set the null surface $t=t_{0,+}$ as the initial surface
(In the later discussions, we omit the subscript ``$+$").
Note that for $t_0\ll \ell$,
\beq
t_0 \approx -\frac{\ell^2}{8{\cal M}}.
\eeq

By introducing the new coordinate $T:=t-t_0$,
we obtain
\beq
\Delta(T)
&=&\frac{T}{4\ell^2}
 \big(T^3+4t_0 T^2+(2\ell^2+6t_0^2)T+4(t_0^3+\ell^2t_0+2{\cal M}\ell^2))
\nonumber\\
&=&\frac{T}{4\ell^2}
 \big(T^3+4t_0 T^2+(2\ell^2+6t_0^2)T+3t_0^3+2\ell^2t_0-\frac{\ell^4}{t_0}\big).
\eeq
In the new coordinate system, the spacetime metric can be written as
\beq
ds^2=-n(T)^2 dT^2
+a(T)^2\Big(\big(dx^1\big)^2 +\sin^2 x^1 \big(dx^2\big)^2 \Big)
+c(T)^2\Big(dx^3+\cos x^1 dx^2\Big)^2\,.
\eeq
To compare the time evolutions of two independent scale factors, it is useful to rewrite the metric
\beq
ds^2=-n(T)^2 dT^2
+e^{2\alpha}
\Big[
e^{2\beta(T)}\Big(\big(dx^1\big)^2 +\sin^2 x^1 \big(dx^2\big)^2 \Big)
+e^{-4\beta(T)}\Big(dx^3+\cos x^1 dx^2\Big)^2
\Big]
\,,
\eeq
where
$\alpha=(1/3)\ln (a^2c)$ and $\beta=(1/3)\ln\big(a/c\big)$.
The dynamics of Bianchi IX Universe and in particular
the Taub-NUT de Sitter spacetime can be understood
in terms of an analogy with the classical mechanics,
as is discussed in Appendix A.
See \cite{dechant} for the other Bianchi IX
solutions which are coupled to a scalar field.

\section{Scalar field in the Bianchi I model}

We consider a massless, minimally coupled scalar field
propagating on the background spacetime discussed in the previous section
with action
\beq
S_{\phi}=-\frac{1}{2}
\int d^4 x\sqrt{-g}
\Big(g^{\mu\nu}\partial_{\mu}\phi\partial_{\nu}\phi
\Big).
\eeq
Note that this scalar field is not the inflaton and hence does not contribute to the background dynamics. We focus on the quantization procedure and
the evolution in the anisotropic stage. The effects of a scalar field onto the geometry of the anisotropic Universe were discussed in \cite{ms}.

\subsection{Equation of motion}

Explicitly, the scalar action is written as
\beq
S_{\phi}=\frac{1}{2}\int d\tau d^3x
e^{3\alpha}
\Big(\phi_{,\tau}^2
-e^{-2\alpha} h^{ij}
\partial_i \phi
\partial_j \phi
\Big).
\eeq
Introducing the Fourier decomposition for the mode function
\beq
\phi=\int \frac{d^3k}{(2\pi)^{3/2}}
     \phi_{\gamma}e^{i {\bf k}{\bf x}}\,,
\eeq
where $\gamma=(k_1,k_2,k_3)$ denotes
{the mutually orthogonal components
of the comoving momentum},
the action can be reduced to
\beq
S_{\phi}=\frac{1}{2}\int d^3k d \eta
\Big[\Big(\frac{d\chi_{\gamma}}{d\eta}\Big)^2
-{\omega}_{\gamma}^2{ \chi}_{\gamma}^2\Big],
\eeq
where
$\chi_{\gamma}=e^\alpha\phi_{\gamma}$
and $d\eta=e^{-\alpha}d\tau$, which corresponds to the conformal time
in the isotropic case.
The frequency squared is given by
\beq
 \omega_{\gamma}(t)^2&=&
- 2H^2 e^{2\alpha}+e^{-4\alpha}\left[H^2 +\Omega^2(t)\right],
\eeq
where $H= \sqrt{\frac{\Lambda}3}$ is the Hubble parameter
for a large $\tau$
and the momentum dependent part is
\beq
\Omega^2(t) &=&
\frac{\Big(1+\sqrt{1+e^{6\alpha}}\Big)^{\frac23(m+\sqrt{3(1-m^2)})}}
{e^{2(m-2+\sqrt{3(1-m^2)})\alpha}} k_1^2
+ \frac{\Big(1+\sqrt{1+e^{6\alpha}}\Big)^{\frac23(m-\sqrt{3(1-m^2)})}}
      {e^{2(m-2-\sqrt{3(1-m^2)})\alpha}} k_2^2
+\frac{e^{4(m+1)\alpha}}
      {\big(1+\sqrt{1+e^{6\alpha}}\big)^{\frac{4m}3}} k_3^2.
\eeq
The equation of motion
for each Fourier mode becomes
\beq
\frac{d^2\chi_{\gamma}}{d\eta^2}
+\omega_{\gamma}^2\chi_{\gamma}
=0\,.
\eeq

In the following discussions, we will focus on
the background solutions which have the planar symmetry,
$m=-1$ or $m=+1$.
For the branch $m=-1$, in the limit $a\to 0$,
\beq
\omega_{\gamma}^2= \frac{H^2+2^{4/3}k_3^2}{a^4}
+\Big(-2 H^2 +\frac{3k^2+k_3^2} {3 \times 2^{2/3}} \Big)  e^{2\alpha}
+\cdots\,,
\eeq
where $k^2:= k_1^2+k_2^2+k_3^2=k_\perp^2+k_3^2$.
In the later time limit $e^\alpha \to \infty$,
\beq \label{large a}
\omega_{\gamma}^2=-2 H^2 e^{2\alpha} +k^2
-\frac{k^2 -3k_3^2}{3e^{3\alpha}}
+\frac{H^2}{e^{4\alpha}}
+O(e^{-5\alpha})\,,
\eeq
which approaches the one in the ordinary de Sitter universe for large scale factor $e^\alpha$.
The case of the branch of $m=+1$ will be discussed in subsection II. G.

\subsection{Quantization and power spectrum}

The canonical quantization of the scalar field is done
in the standard manner:
\beq
\phi= \int d^3k
    \Big(u_{\bf k} a_{\bf k}
+    u^{\ast}_{\bf k} a^{\dagger}_{\bf k}
\Big),
\eeq
where $\big[ a_{\bf k_1},a^{\dagger}_{\bf k_2}\big]
=\delta({\bf k_1}-{\bf k_2})$ (others are zero)
and $u_{\bf k}= e^{i{\bf k}{\bf x}}\phi_{\bf k}/(2\pi)^{3/2}$,
and the mode function in the time-direction can be normalized as
$
\phi_{\bf k}\partial_t \phi^{\ast}_{\bf k}
-\big(\partial_t \phi_{\bf k} \big)\phi^{\ast}_{\bf k}
=i/e^{3\alpha}.
$
The power spectrum is defined by
\beq
\big\langle 0| \phi^2 |0\big \rangle
:=\int d\ln k\, \int\frac{d\theta_{\bf k}}{2} P_{\phi},\quad
P_{\phi}=\frac{k^3}{2\pi^2} \big|\phi_{\bf k} \big|^2  .
\label{ps}
\eeq
Note that in contrast to the case of the standard inflation
the direction dependence would be included in the power spectrum.
The vacuum is chosen at the initial anisotropic era: $\tau \to 0-$.
Usually, in a quantum harmonic oscillator system,
an {\it adiabatic} process implies the
one where the potential changes slowly enough compared to its size,
and 
the time evolution can be obtained from the zero-th order
WKB approximation.
In this paper , we follow this definition for the term {\it adiabatic}. 
In the standard inflationary models,
an adiabatic vacuum is also defined in the same way. 
Note that an {\it isentropic} process implies 
$(adiabatic) + (reversible)$ process
and therefore is slightly different from our {\it adiabatic} one.  
In the anisotropic spacetime
an adiabatic vacuum state can be found only
in the special solutions of the anisotropic Universe,
which are regular in the initial times.
We call our adiabatic vacuum {\it an anisotropic vacuum}.

\subsection{WKB solution for $m=-1$}

From now on, we will focus on the branch of $m=-1$
{except in the subsection II F}.
By changing the coordinate $dt=d \tau /e^{3 \alpha}$, which leads to
$$
e^{\alpha} = \sinh^{1/3} (3 H\tau) = \frac{1}{\sinh^{1/3}(-3Ht)}\,,
$$
one obtains the equation of motion
in the form of a time-dependent oscillator
\beq \label{eom}
\Big(
\frac{d^2}{dt^2}
+\Omega(t)^2
\Big)\phi
=0\,.
\eeq
When $\tau$ changes from $0+$ to $+\infty$,
$t$ changes from $-\infty$ to $0-$.
The frequency squared becomes
\beq
\Omega^2(t)
:=\frac{2^{4/3}\big(k_\perp^2e^{6H t}+k_3^2\big)}
      {(1-e^{6H t})^{4/3}}=\frac{2^{4/3} k^2}{x^{4/3}}(1- r_\perp^2 x),
\eeq
where $r_\perp:=k_\perp/k$ and $x(t) =1- e^{6Ht}= e^{-6\alpha}(\sqrt{e^{6\alpha}+1}-1) $ monotonically varies from one to zero as time $t$ increases from negative infinity to zero.

\begin{figure}[tbph]
\begin{center}
\begin{tabular}{ll}
\includegraphics[width=.5\linewidth,origin=tl]{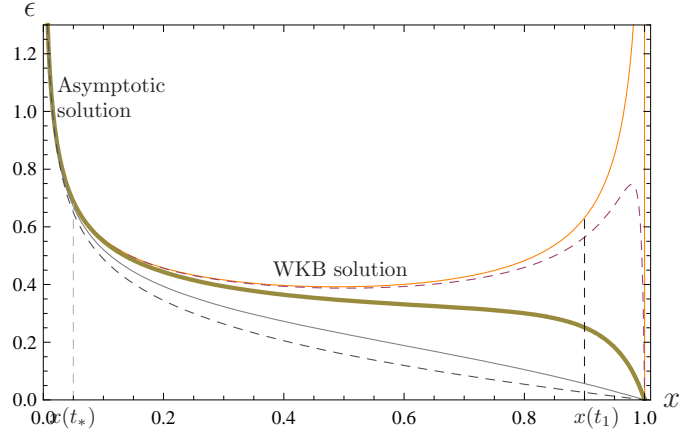}
\end{tabular}
\end{center}
\caption{ The behavior of $\epsilon(t)$ with respect to $x$. Here, $r_\perp^2 = 0.999, 0.99, .9, .5,$ and $ .1$ respectively from the top and we choose $H/k= 0.05$. The WKB approximation is valid if $\epsilon(t)< 1$.
The matching time $t_\ast$ is almost independent of $k$.
} \label{fig:epsilon}
\end{figure}

The WKB solution is given by
\beq
\phi_{\rm WKB}=\frac{1}{\sqrt{2\tilde \Omega(t)}} \left\{
\exp\Big[-i\int_{t_0}^t dt'\tilde \Omega (t') +i \psi\Big]
\right\},
\eeq
where $\psi$ is a phase factor and $H |t_0|\gg 1$ and $\tilde \Omega$ satisfies
the nonlinear equation
\beq
\tilde \Omega (t)^2= \Omega (t)^2
   -\frac{1}{2}
    \Big(\frac{\tilde \Omega_{,tt}}{\tilde \Omega}
        -\frac{3\tilde \Omega_{,t}^2}{2\tilde \Omega^2}
    \Big).
\eeq
The WKB wavefunction is valid up to the order of the correction term if
\begin{eqnarray}\label{WKBcondition}
\epsilon(t):= \left|\frac{\frac{d\Omega^2(t)}{dt}}{\Omega^3(t)}\right|= \frac{H}{k}\frac{ 1-x(t) }{(x/2)^{1/3}\big(1- r_\perp^2 x(t)\big)^{1/2}}\left(\frac{3}{1- r_\perp^2x(t)}+
    1\right) \ll 1 ,
\end{eqnarray}
where $\epsilon$ play the role of the adiabatic parameter and is plotted in Fig.~\ref{fig:epsilon} for several different values of $ r_\perp$.
At $x=1$, all the curves converge to zero
and therefore WKB approximation is always valid at the beginning.
Thus, it is possible to define an anisotropic vacuum.
In the case of $m=-1$,
the initial spacetime metric can be approximately written as
\beq
ds^2\approx -d\tau^2+ d(x^1)^2+d(x^2)^2+\tau^2 d(x^3)^2,
\label{mm}
\eeq
where unimportant constants are absorbed
by the rescaling of $x^i$.
It is straightforward to see that this metric
represents (a patch of) the Minkowski spacetime:
\beq
ds^2=-dU^2+ d(x^1)^2+d(x^2)^2+ dR^2,
\eeq
by the coordinate transformation $R=\tau \sinh x^3$ and $U=\tau \cosh x^3$.
Thus, it is possible to
find the positive frequency modes
and the associated vacuum state,
for an observer moving toward
the increasing $\tau$-direction along a constant $x^3$
curve on the $(U,R)$-plane.
This corresponds to our choice of the vacuum at the beginning time.

Near $x\sim 1$, the condition is always valid for all momentum modes except for the mode with $r_\perp =1$.
The condition is also valid for high momentum modes if the scale factor $ e^\alpha$ satisfies $e^\alpha \ll \frac{k}{H}$ {\it i.e.,} $-Ht \gg \left(\frac{H}{k}\right)^3$.
On the other hand, the large $ e^\alpha$ approximation~(\ref{large a}) is valid if
$e^\alpha \gg 1 $.
For high frequency modes, if we choose an intermediate time $t_\ast$ satisfying
\begin{eqnarray} \label{cond:adiabatic}
1\ll e^{\alpha(t_\ast)} \ll \frac{k}{H},
\end{eqnarray}
and choose the WKB wavefunction for $t<t_\ast$ and the large $e^\alpha$
approximation,
for $t>t_\ast$ the approximation will be good and in fact its accuracy is
almost independent of the choice of the intermediate time $t_\ast$ as long as it satisfies Eq.~(\ref{cond:adiabatic}).
Therefore, we choose
\begin{eqnarray} \label{t_ast}
e^{\alpha(t_\ast)} = \sqrt{\frac{k}H}
\end{eqnarray}
and the adiabatic parameter $\epsilon_{\ast}$ at $t=t_{\ast}$, is order of $O\big(H^{1/2}/k^{1/2}\big)$.
For modes with $k/H \sim 1$ or $k< H$, the approximation is not valid.
The WKB wavefunction is expanded up to {an}
enough
adiabatic order, to validate our matching scheme with the solutions
in the quasi-de Sitter region.
The assumption {of} $k\gg H$ is justified if we assume a slow rolling inflation for our universe.
Note that however, as we will discuss later,
for a given class of inflation models,
the modes of $k\gtrsim H$
are observable in some cases.

Near $x\sim 0$, 
the WKB solution at the zeroth adiabatic order takes the form:
\begin{eqnarray}
\phi_{\rm WKB} = \frac{1}{\sqrt{2\Omega(t)}} \exp\left[i\frac{2^{2/3}k}{6H}\left(\psi_k+3 x^{1/3}+\frac{3}{4}\big(1-\frac{r_\perp^2}2\big)x^{4/3}+\cdots \right) \right].
\end{eqnarray}
Note that the next order correction at $e^{\alpha(t_\ast)} = \sqrt{k/H}$ is $O(H/k)$ since $x_\ast:=x(t_\ast)= 2(H/k)^{3/2}(1-(H/k)^{3/2}+\cdots)$.
The value of the frequency $\Omega_\ast:=\Omega_{k}(t_\ast)$ is
 \beq
 \Omega_\ast = \frac{k^2}{H} \left[1+\left(\frac23-r_\perp^2\right)
\left(\frac{H}k\right)^{3/2}+\cdots \right].
\label{omegaast}
 \eeq
The integral on the exponent is evaluated as
\beq
&&k\int^t_{t_0}
dt \frac{2^{2/3}\sqrt{1-r_\perp^2(1- e^{6Ht})}}{(1-e^{6Ht})^{2/3}}
=-\frac{2^{2/3}k}{6H} \int_{1-\epsilon}^{x} \frac{dx}{1-x}
\frac{(1-r_\perp^2 x)^{1/2}}{x^{2/3}},
\eeq
where
$\epsilon:= e^{6Ht_0}$ is some cut-off scale at an initial time.
The integration can be executed explicitly to give the Appell hypergeometric function of two variables,
$$
 \int \frac{dx}{1-x}\frac{(1-r_\perp^2 x)^{1/2}}{x^{2/3}}
  =\frac{2r_\perp^{4/3}(1-r_\perp^2 x)^{3/2}}{3(1-r_\perp^2)} F_1(\frac{3}{2},
\frac{2}{3},1,\frac{5}2,1-r_\perp^2 x,\frac{1-r_\perp^2 x}{1-r_\perp^2}) .
$$

\subsection{Matching}

In the (quasi-) de Sitter region, $(-t)\ll H^{-1}$,   
the equation of motion~(\ref{eom}) becomes
$$
(\frac{d^2}{dt^2}+ \frac{k^2}{(-3Ht)^{4/3}}) \phi = 0,
$$
whose solution becomes
\beq \label{phi:fin}
\phi_{\rm fin} =A_+ \phi_+(t) +A_-\phi_-(t)\,,
\eeq
where the positive and negative frequency modes are given by
\beq \label{phi:+-}
\phi_+(t):=\Big(-1+ \frac{i k}{H}(-3Ht)^{1/3}\Big)
e^{\frac{i k}{H}(-3Ht)^{1/3}},\quad
\phi_-(t):=\Big(-1-\frac{i k}{H}(-3Ht)^{1/3}\Big)
e^{-\frac{i k}{H}(-3Ht)^{1/3}}.
\eeq
This approximation is valid for $ e^\alpha \gg 1$.
The WKB solution is now matched to the one in the quasi-de Sitter region, at $t=t_{\ast}$.
Now, we should determine the adiabatic order where the matching is done.
As an example, we post the frequency $\Omega$ at $t_\ast$ by using $(-3Ht_{\ast})=\sinh^{-1}(e^{-3\alpha(t_\ast)})=\sinh^{-1}(\big(H/k\big)^{3/2})$ and
Eq. (\ref{omegaast}):
\begin{eqnarray}
&& \Omega_\ast = k \left(\frac{k}{H}\right)
  \left [1+2(\frac{2}3-r_\perp^2) \left(\frac{H}{k}\right)^{3/2} +\cdots \right] .
\end{eqnarray}
As seen here, the first direction dependence appears at the third order in $\epsilon^3$
and we need the WKB approximation up to the fourth adiabatic order.
Then, the WKB frequency is given by
\beq
\tilde \Omega (t_\ast)^2= \left[\Omega (t)^2
   -\frac{1}{2}
    \Big(\frac{\Omega_{,tt}}{ \Omega}
        -\frac{3 \Omega_{,t}^2}{2 \Omega^2}
    \Big)
+O(\epsilon^4)\right]_{t_\ast}
=\Omega_\ast^2\Big(1-\frac{2H}{k}\Big)
+O(\epsilon_\ast^4)
.
\eeq

The matching condition is given by
\beq
&&
A_+ \phi_+(t_{\ast})
+  A_- \phi_-(t_{\ast})=\left(1+\frac{H}{2k}\right)\frac{\Phi_{\ast}}{\sqrt{2\Omega_{\ast}}} ,
\nonumber\\
&&
A_+ \phi_+{}'(t_{\ast})
+  A_- \phi_-{}'(t_{\ast})=
-i\left[(1-\frac{H}{2k})- i \sqrt{\frac{H}{k}}\left(1-\frac{H}{2k}\right)  \right]\sqrt{\frac{\Omega_{\ast}}{2}}
\Phi_{\ast}.
\eeq
Now the solution is exact up to the adiabatic order of $\epsilon^4$.
The phase factor is given by
\beq
\Phi_{\ast}
={\rm exp}
\Big[
-i \int_{t_0}^{t_{\ast}}\tilde \Omega dt+i \psi
\Big].
\eeq
It is straightforward to confirm that the coefficients $A_+$
and $A_-$ satisfy the normalization condition
\beq
\big|A_{+}\big|^2-\big|A_{-}\big|^2
=\frac{H^2}{2k^3}\,.
\eeq
As the reference,
in the standard isotropic inflation, the normalized mode function is given by
\beq
\phi_{\rm isotropic}
=\frac{i H}{\sqrt{2k^3}}
\Big(-1+i\frac{k}H (-3Ht)^{1/3}\Big)
e^{i\frac{k}H(-3Ht)^{1/3}}.\label{iso}
\eeq

\subsection{Power spectrum}

In the limit of the later times $t\to 0-$, the wavefunction behaves as
\beq
 \phi_{\rm fin}\Big|_{t\to 0}
&=&-(A_+ + A_-). \nn
\eeq
Explicitly,
{the wavefunction} in the later time to the order $\epsilon_\ast^3$ becomes
\begin{eqnarray*}
\phi_{\rm fin} &=& -\frac{\Phi_\ast}{\sqrt{2k}} \left(\frac{H}{k}\right)\left(1-\frac H{2k} \right)
    \left(1-\frac12\left(\frac23-r_\perp^2\right)\left(\frac{H}k\right)^{3/2}  \right)
    \times \nn\\
   && \left\{\left(\sin\sqrt{\frac{k}H} +\sqrt{\frac H k} \cos\sqrt{\frac{k}H}\right)
    +i \left(1+\left(\frac23-r_\perp^2\right)\left(\frac{H}k\right)^{3/2}  \right) \left[\cos \sqrt{\frac{k}H}-\sqrt{\frac H{k}}\sin\sqrt{\frac{k}H}   \right]\right\}.
\end{eqnarray*}

Then, the power spectrum Eq. (\ref{ps}) in terms of the adiabatic order $O((H/k)^{3/2})$ is given by
\beq
P_{\phi}
&=&\frac{H^2}{4\pi^2}
\left\{
1-\frac{H}{k} +\left(\left(\frac{2}{3}-r_\perp^2\right)\left(\frac{H}{k}\right)^{3/2}
-\frac{2H}{k}\right)\cos 2\sqrt{\frac{k}H} +O\left(\frac{H}k\right)^2 \right\}.
\eeq
As a reference, we present the power spectrum from the standard inflation
obtained from Eq. (\ref{ps}) and (\ref{iso}):
\beq
P_{\phi}^{(0)}=\frac{H^2}{4\pi^2}.\nn
\eeq
There are two modifications:
One is an oscillatory behavior of the power spectrum, which
seems to be observed in the CMB angular power spectrum
around the first acoustic peak $\ell=10\sim 100$,
and the other is an overall change of the amplitude given by $(1-H/k)$.
In the above expression by taking $k\to H$, the power spectrum seems to vanish.
This would be consistent with the suppression of the angular power spectrum of CMB fluctuations on large scales if the anisotropy of the present Hubble horizon scale corresponds to the wavenumber of the order $k \gtrsim H$.
For larger $k$, it approaches the scale invariant form
with a small oscillatory behavior.
The direction dependence appears only on the oscillatory term.

As discussed above,
our WKB approximation is valid only for modes with $k/H\gg 1$.
Indeed, one may ignore the contributions of modes such as $k\sim H$
or $k<H$.
One may not observe modes whose wavelengthes are longer than the present Hubble scale.
Thus, there is the critical wavenumber given by $k_{\rm crit}=a_0 H_0$, where $H_0^{-1}\sim 10^{28}cm$
is {present day's Hubble scale:
one observes the modes of $k>k_{\rm crit}$}.
If it is possible to take $k_{\rm crit}>H$, our WKB approximation is
justified for all the observable modes.
On the other hand, $a_0=\big(T_{\rm R}/T_0\big) a_{\rm f}$, where
$T_{\rm R}$, $T_{0}\approx 2.7K$ and $a_{\rm f}$ are the reheating temperature,
the CMB temperature at present and the value of scale factor
at the end of inflation, {respectively}.
One finds the relation
\beq
k_{\rm crit}=a_0 H_0=\frac{a_f T_{\rm R}}{T_0}H_0
    =e^N\frac{T_{\rm R}}{T_0}\frac{H_0}{H} H
    \simeq
    A H,\quad
A:= e^N\frac{T_{\rm R}}{T_0}\frac{V_{\rm DE}^{1/2}}{V_{\rm inf}^{1/2}},
\eeq
where $N=\ln (a_{\rm f}/a_{\rm i})=\ln a_{\rm f}$ is the number of $e$-folding of inflation,
$V_{\rm inf}$ and $V_{\rm DE}\sim (10^{-3}{\rm eV})^4$ are the values of inflation and dark energy potentials.
Note that
$H\approx \sqrt{(8\pi/3M_p^2) V_{\rm inf}}$
is the Hubble constant
during inflation and
$H_0\simeq \sqrt{(8\pi/3M_p^2) V_{\rm DE}}$ is that of the present Universe.
One can estimate the value of $A$ as
$$
A\simeq e^{N-64} \left(\frac{T_{\rm R}}{10^{14} {\rm GeV}}\right)
                 \left(\frac{10^{16} {\rm GeV}}{V_{\rm inf}^{1/4}}\right)^2\,.
$$
Thus, for inflation with GUT or a lower energy scale
and if $e$-folding is around $N\gtrsim 64$,
it may be possible to realize $A\gtrsim 1$, leading to
$k_{\rm crit}>H$.
On the other hand,
for sufficiently long duration of inflation, say $N\gg 64$,
the observable modes are only those with $k \gg H$ since
$A\gg 1$,
and therefore it seems to be very hard to find
any deviation from the scale-invariant spectrum.
In other words, to find the information about
the preinflationary anisotropy,
the number of $e$-folding must be tuned.

\subsection{Planar modes, $r_{\perp}\lesssim 1$ }

As seen in the case of $r_\perp^2=0.999$ in Fig.~\ref{fig:epsilon},
{namely}
for modes on a plane, there appears a region where the WKB approximation
may not {be} valid near $x\sim 1$.
We divide the time into three separate regions divided by the times $t_{1}$ and $t_{\ast}$.
In the region $t_1<t< t_{\ast}$ the WKB approximation is valid.
For other two regions, we may find an appropriate approximate solution.
Note that in the case of $r_{\perp}=1$ exactly,
the adiabaticity parameter {diverges}
in the limit of $t\to -\infty$
and there is no anisotropic vacuum state.

The frequency squared for $x\simeq 1$ becomes
$$
\Omega^2 = 2^{4/3}(k_3^2+k_\perp^2 e^{6Ht})+O(k^2e^{12Ht}), \quad \mbox{for } Ht\ll -1 .
$$
The corresponding solution for Eq.~(\ref{eom}) is
\begin{eqnarray} \label{phi1}
\phi_1= \sqrt{\frac{\pi }{6H \sinh(\pi q_3)}} \, J_{-iq_3}(q_\perp e^{3Ht}), \quad  \mbox{for } Ht \ll -1,
\end{eqnarray}
where
$$
q_3 = \frac{2^{2/3} |k_3|}{3H}, \quad q_\perp = \frac{2^{2/3} k_\perp}{3H}.
$$
{
Here we chose the solution so that it becomes an incoming wave form}:
\beq
\phi_{1}=\frac{1}{\sqrt{2 \cdot 2^{2/3} |k_3|}} \exp
\Big\{- 2^{2/3}\,i |k_3| t+i\psi \Big\}, \quad  \mbox{for } t\to -\infty, \nn
\eeq
where the initial phase becomes
$e^{i \psi} =\left[ \Gamma(1+i q_3)/\Gamma(1-i q_3)\right]^{1/2} \left(\frac{q_\perp}2\right)^{iq_3}$.
We choose the momentum dependent cutoff time $t_1$ so that both of the solution~(\ref{phi1}) and the WKB solution are equally valid.
The WKB solution is valid if $\epsilon < 1$.
For $x\sim 1$, we check the validity of the WKB approximation,
\beq
\epsilon\approx \frac{3\cdot 2^{1/3}H}{k}
    \frac{e^{6Ht}}{(r_3^2 +e^{6Ht})^{3/2}} \simeq \frac{3\cdot 2^{1/3}H}{k}\frac{1}{e^{3Ht}} \,,\nn
\eeq
where in the second equality we assume $e^{3Ht} \gg r_3$ since we want $t_\ast$ to be placed over the first peak  in $t$
as seen in Fig.~\ref{fig:epsilon}.
For the WKB approximation to be valid, we should choose the time so that $e^{3Ht_1} > \frac{3\cdot 2^{1/3} H}{k}$.
For example, we may choose the time $t_1$ satisfying
\beq
e^{3Ht_1} = \frac{3}{2^{2/3}}\sqrt{\frac{H}{k}},
\eeq
where both of the WKB approximation and the solution~(\ref{phi1}) are equally good.
Note that the time $t_1$ is direction independent.
At time $t_1$, the assumption $e^{3Ht}\gg r_3$ constrains the orthogonal momentum value to
\begin{eqnarray} \label{q3}
q_3 \ll \sqrt{\frac{k}{H}}.
\end{eqnarray}
For  planar high-momentum modes, the argument of the Bessel function, $q_\perp e^{3Ht_1}=\sqrt{\frac{k_\perp^2}{kH}}\simeq \sqrt{\frac{k}{H}}$, is very large.
Therefore at $t_1$, we may employ the asymptotic form for the Bessel function:
\begin{eqnarray}
\lim_{x\to \infty}J_{ia}(x) = \sqrt{\frac{2}{\pi x}} \cos\left(x-\frac{\pi}4-\frac{ia\pi}{2}\right) .
\end{eqnarray}
During $t_1< t< t_\ast$, we use the WKB approximation where $t_\ast$ is given in Eq.~(\ref{t_ast}).
The WKB solution is
\beq \label{phi2}
\phi_2=
\frac{B_+}{\sqrt{2\Omega}}
\exp
\Big\{-i \int_{t_{1}}^{t} dt' \Omega(t')\Big\}
+\frac{B_-}{\sqrt{2\Omega}}
\exp
\Big\{i \int_{t_{1}}^{t} dt' \Omega(t')\Big\}.
\eeq
For $t\gg t_{\ast}$, the WKB approximation is not valid.
However, for $t>t_\ast$ the asymptotic solution~(\ref{phi:fin}) is valid.

At $t_{1}$, to the zeroth order in $H/k$, the matching condition becomes
\begin{eqnarray}
B_+ &=& \sqrt{\frac{\Omega_1}{2}} \left[\phi_1(t_1)+i\frac{\phi_1'(t_1)}{\Omega_1}\right]
 \simeq  \frac{i}{\sqrt{2\sinh(\pi q_3)}}\left[\left(1+\frac{Hq_3^2}{k}\right)^{-1/4}\cos \psi  -i \left(1+\frac{Hq_3^2}{k}\right)^{1/4}\sin \psi  \right]
    , \nn  \\
 B_- &=&  \sqrt{\frac{\Omega_1}{2}} \left[\phi_1(t_1)-i\frac{\phi_1'(t_1)}{\Omega_1}\right]
  \simeq  \frac{-i}{\sqrt{2\sinh(\pi q_3)}} \left[\left(1+\frac{Hq_3^2}{k}\right)^{-1/4}\cos \psi  +i   \left(1+\frac{Hq_3^2}{k}\right)^{1/4}\sin \psi  \right]  , \nn
\end{eqnarray}
where $\Omega_1^2 \simeq 9\left(\frac{H}{k}k_\perp^2 + \frac{2^{4/3}}{9} k_3^2\right) = 9H^2\frac{k_\perp^2}{H k}\left( 1+\frac{H k}{k_\perp^2}\frac{Hq_3^2}{k}\right)$ and $\psi =\frac{i \pi q_3}{2} + \frac{\pi}4+ \frac{k_\perp}{\sqrt{kH}}$.
Since $q_3^2 \ll k_\perp^2/(kH)$ as in Eq.~(\ref{q3}), the coefficient $B_\pm$ becomes {of the zeroth order},
\begin{eqnarray} \label{B+-}
B_+ \simeq \frac{1}{\sqrt{1-e^{-2\pi q_3}}}
\times e^{i (\frac{\pi}4-\sqrt{\frac{k_\perp^2}{k H}})} , \quad
B_- \simeq \frac{e^{-\pi q_3}}{\sqrt{1-e^{-2\pi q_3}}}
\times e^{-i (\frac{\pi}4-\sqrt{\frac{k_\perp^2}{k H}})} .
\end{eqnarray}

For $t_\ast<t<0$, the isotropic solution is valid and
\beq
\phi_3=C_+ \phi_+(t) + C_- \phi_-(t)\,.
\eeq
The primordial power spectrum is proportional to the square of the size of $C_++C_-$.{
Matching this solution
with the WKB solution~(\ref{phi2}) at the boundary $t_\ast$,} we have
$$
C_++C_- = \frac{H^2}{2ik^3} \left[\phi_2(t_\ast) (\phi_-'(t_\ast)-\phi_+'(t_\ast)) -\phi_2'(t_\ast) (\phi_-(t_\ast)-\phi_+(t_\ast))\right],
$$
Using Eqs.~(\ref{B+-}) and (\ref{phi:+-}), we get, %
in the limit $q_3^2 \ll k_\perp^2/(kH)$,
\begin{eqnarray}
C_++C_- &\simeq &  -\frac{H}{\sqrt{2(1-e^{-2\pi q_3})}k^{3/2}}
    \left\{- (1+e^{-\pi q_3})
    \sin(\frac\pi 4-\Phi(k) )
    +i (1-e^{-\pi q_3})
    \cos(\frac\pi 4-\Phi(k))
   \right\},
\end{eqnarray}
where $\Phi(k)=\sqrt{\frac{k_\perp^2}{kH}}+ \int_{t_1}^{t_\ast} dt' \Omega(t') +\sqrt{\frac{k}H}$ and we ignore terms of $O(\sqrt{H/k})$.

The primordial power spectrum to the zeroth order is given by
\beq
P_{\phi}&=&
\frac{k^3}{2\pi^2}
\Big[
 \Big\{ {\rm Re}\Big(C_+ + C_-\Big)\Big\}^2
+\Big\{ {\rm Im}\Big(C_+ + C_-\Big)\Big\}^2
\Big]\, \\
&=& \frac{H^2(1-e^{-\pi q_3})}{4 \pi^2} \left[1+\frac12 {\rm cosech}^2\frac{\pi q_3}{2}\left(1- \sin(2\Phi(k)) \right)
 \right] . \nn
\eeq
If one reduces $q_3$ keeping $k$ constant, the direction dependence of the power spectrum becomes un-negligible at $q_3 \sim 2/\pi$ or $ k_3 \sim \frac{3\cdot 2^{5/8}}{\pi} H \ll k $.
Note that this result is quite different from the general expectation~\cite{anomaly} that the direction dependent term would be (approximately) scale invariant and second order in $k_3$.
The phase factor $\Phi(k)$ is dominated by $\int_{t_1}^{t_\ast} dt' \Omega(t') $ and is becomes
$$
2\Phi(k) =2\sqrt{\frac{k_\perp^2}{kH}}+2\int_{t_1}^{t_\ast} dt' \Omega(t')  +2\sqrt{\frac{k}H}\simeq \frac{2^{2/3}\sqrt{\pi}\Gamma(1/3)}{3\Gamma(5/6)} \frac{k}{H}.
$$

\subsection{The case of $m= +1$}

In this subsection, we briefly discuss the case of
the other solution with a planar symmetry, $m=+1$.
We will see
that the WKB approximation can not be applied
at the beginning
and there is no anisotropic vacuum.
In this case, it is appropriate to change the variables
as $\phi=a^{-1/2} \Phi$ and $dt=a^{-1}du$,
to rewrite the equation of motion
into the form of a time-dependent harmonic oscillator:
\beq
\Big[
\frac{d^2}{du^2}
+\Xi(u)^2
\Big]\Phi(u)=0\,,
\eeq
where
\beq
\Xi(u)^2:=\frac{a'{}^2}{4a^2}
-\frac{a''}{2a}
+a^{-2}\Omega^2
=\frac{a'{}^2}{4a^2}
-\frac{a''}{2a}
+k^2\frac{r_{\perp}^2(1+\sqrt{1+a(u)^6}) + a(u)^6 (1-r_{\perp}^2)}
         {(1+\sqrt{1+a(u)^6})^{4/3}}\,,
\eeq
and the prime denotes the derivative with respect to $u$.
The coordinate $u$ is defined
in the range $0\leq u<u_0:=\sqrt{\pi}\Gamma(1/3) /(3H\Gamma(5/6))
\approx  1.402  H^{-1}$.
In order to check the validity of the WKB approximation,
one needs another coordinate transformation such that
$H v=\ln (u/u_0)$ and $\Phi=e^{vH/2}\Psi$,
where $v$ runs from $-\infty$ to zero,
and then the equation of motion becomes
\beq
\Big[\frac{d^2}{dv^2}+\xi(v)^2\Big]\Psi=0\,,
\eeq
where
\beq
\xi(v)^2:= H^2 u_0^2 e^{2v} \Xi(u(v))^2-\frac{H^2}{4}.
\eeq
The validity of WKB approximation is
determined by the parameter
$
\epsilon_v:=\big|\big(d\xi^2/dv\big)/\xi^3\big|.
$
For $t\to -\infty$, $u\to 0$ and $v\to -\infty$, then $a\approx H u$,
and one can evaluate $\xi^2\approx k^2 r_{\perp}^2 H^2 u_0^2 e^{2H v} $
and
$
\epsilon_v\approx 2e^{-H v}/k_{\perp}u_0\,.
$
Therefore, the adiabaticity parameter exceeds
{unity} for $v \ll - (1/H) \ln \big(k_{\perp}u_0 \big)$.
In such a case, one cannot find a well-defined adiabatic vacuum state.

\section{Scalar field in the Bianchi IX Universe}

In this section, we investigate
the primordial spectrum of a massless and
minimally coupled scalar field
in the Bianchi IX Universe.
Such an investigation
will give a deep insight about
how generic the results obtained
in the case of the Bianchi I model are.
One reason is that
recent WMAP data
seem to favor a spatially closed Universe
\footnote{
Simultaneous constraints on
the spatial curvature and
the (constant) equation of state parameter of dark energy
by the five-year WMAP data
seem to favor a positive curvature,
although the flat Universe is also
consistent at 95 percent confidence level \cite{komatsu}.},
and the Bianchi IX Universe exactly has that geometry.
The other important reason is that,
similarly to the case of the Bianchi I model,
in the Bianchi IX model there is an exact analytic solution,
the Taub-NUT de Sitter spacetime whose metric is given
by Eq. (\ref{met2}).
The spacetime geometry is a Taub-NUT spacetime
in the initially anisotropic era, and
approaches de Sitter solution in the later time.
Our discussion will come along the same line in the case of the Bianchi I.
Note that as discussed in Sec. II,
{we restrict the value of ${\cal M}$ to be ${\cal M}\geq 2\ell/(3\sqrt{3})$,}
for which $\Delta(t)$ vanishes at $t=t_0<0$ ($T=0$).
For such a choice,
one can naturally choose the initial time to be $T=0$.

\subsection{Scalar field on the Bianchi IX Universe}

The equation of motion of a massless and minimally coupled
scalar field, $\Box \phi=0$,
explicitly reduces to
\beq
-\frac{1}{n^2}
\Big[
 \partial_T^2
+\Big(3\dot{\alpha}-\frac{\dot{n}}{n}\Big)
\partial_T
\Big]\phi
+\frac{1}{e^{2\alpha+2\beta}}
\Big[
-L^2
+\Big(-1+e^{6\beta}\Big)
\Big(-L_3^2\Big)
\Big]\phi=0,
\eeq
where the angular momentum operators are defined by
\beq
&&-L^2:=
\Big(\frac{\partial}{\partial x^1}\Big)^2
+\cot x^1 \frac{\partial}{\partial x^1}
+\frac{1}{\sin^2 x^1}
\Big[
 \Big(\frac{\partial}{\partial x^2}\Big)^2
+\Big(\frac{\partial}{\partial x^3}\Big)^2
-2\cos x^1 \frac{\partial^2}
                {\partial x^2 \partial x^3}
\Big],
\quad
-L_3^2:= \Big(\frac{\partial}{\partial x^3}\Big)^2. \nn
\eeq
The general classical solution can be decomposed
by the basis of harmonic functions
$\phi(T,x^i) =\sum_{\gamma}\varphi_{\gamma}(T)Y_{\gamma}(x^1,x^2,x^3)$,
where $\gamma\equiv (J, K, M)$ represents a short-hand notation of
a set of quantum numbers associated with the
spatial directions \cite{bwhu}.
The harmonic function on the three-space can be constructed by
$
Y_{\gamma}(x^1,x^2,x^3):= e^{i K x^3}e^{i M x^2}\Theta_{\gamma}(x^1),
$
where $\Theta_{JKM}$ satisfies the ordinary differential equation
 \beq
\Big[\frac{d^2}{d(x^1)^2}+\cot x^1
     \frac{d}{d(x^1)}
    -\frac{K^2+M^2-2\cos x^1 KM}
          {\sin^2 x^1}
\Big] \Theta_{\gamma}(x^1)
=-J(J+1)\Theta_{\gamma}(x^1)\,. \nn
\eeq
The harmonic function is normalized as
$$
\int dx^1 dx^2 dx^3 \sin x^1
 Y_{\gamma}(x^1,x^2,x^3)Y_{\gamma}^{\ast}(x^1,x^2,x^3)
=\delta_{\gamma, \gamma'}.
$$
The equation in the time-like direction is given by
\beq
0&=&
\Big[\frac{d^2}{dT^2}
+\Big(3\dot{\alpha}-\frac{\dot{n}}{n})\frac{d}{dT}
\Big]\varphi_{\gamma}(\cT)
+\frac{n^2}{e^{2(\alpha-2\beta)}}
\Big[
\frac{J(J+1)-K^2}{e^{6\beta}}
+K^2 \Big]\varphi_{\gamma}(T)
\,.\label{mode}
\eeq

For a given vacuum $|0\rangle$,
the quantization of the scalar field can be done
in the standard way
\beq
\phi(T,x^i)
= \sum_{\gamma}
    \Big(\varphi_{\gamma}(T)Y_{\gamma}(x^i) a_{\gamma}
+    \varphi^{\ast}_{\gamma}(T) Y_{\gamma}^{\ast}(x^i)
a^{\dagger}_{\gamma}
\Big),
\eeq
where the creation and annihilation operators,
$a^{\dagger}_{\gamma}$ and $a_{\gamma}$,
{have} the properties $a_{\gamma}|0\rangle =0$
and $\langle 0| a^{\dagger}_{\gamma}=0$,
and satisfy the
commutation relations
$\big[ a_{\gamma_1},a^{\dagger}_{\gamma_2}\big]
=\delta_{\gamma_1,\gamma_2}$,
and
$\big[ a_{\gamma_1},a_{\gamma_2}\big]
=\big[ a^{\dagger}_{\gamma_1},a^{\dagger}_{\gamma_2}\big]
=0$.
In our later discussions,
a vacuum $|0\rangle$ will be chosen
in order for the initial modes to be adiabatic.
The mode functions satisfy the
Wronskian normalization condition
\beq
\varphi_{\gamma}\partial_T \varphi^{\ast}_{\gamma}
-\partial_T \varphi_{\gamma} \varphi^{\ast}_{\gamma}
=\frac{in}{e^{3\alpha}}.
\eeq
For the case of the discrete spectrum, the power spectrum can be defined
\begin{eqnarray}
\langle 0 | \phi^2 |0 \rangle
=\sum_{\gamma}P_\gamma, \quad \quad P_{\gamma}:=\big|\varphi_{\gamma}\big|^2.
\end{eqnarray}

It is convenient to define the new time coordinate as
\begin{eqnarray}\label{x:t}
dx=ne^{-\alpha+2\beta}\,dT
=\frac{T^2+ 2t_0 T+ t_0^2+\ell^2}{2\ell\Delta}dT
=
\frac{2\big(\cT^2
           +2{\tT_0}{\cal T}
           +{\tT_0}^2
           +1\big)}
{{\cal T}
\big({\cal T}^3+4{\tt}{\cal T}^2
     +\big(6{\tT_0}^2+2\big){\cal T}
     +3{\tT_0}^3+2{\tT_0}-1/\tT_0\big)}d{\cal T},
\end{eqnarray}
and variable as $\varphi=e^{-\alpha-\beta}\chi$,
where $\cT:=T/\ell$ and $\tT_0:=t_0/\ell$ are dimensionless.
Then, the equation of motion is rewritten as
\beq \label{ddchi}
\Big[\frac{d^2}{dx^2}
+\Omega(x)^2
\Big]\chi(x)
=0,
\eeq
where the frequency squared is {given by}
\beq \label{Omega}
\Omega^2(x):=
\Omega_0^2(x)
+\Big(J\big(J+1\big)-K^2\Big)e^{-6\beta}
+K^2.
\eeq
Here the momentum independent part is
\beq
\Omega_0^2(T)
&:=&-\Big(\frac{d\alpha}{dx}+\frac{d\beta}{dx}\Big)^2
-\Big(\frac{d^2\alpha}{dx^2}+\frac{d^2\beta}{dx^2}\Big)
=
-\frac{e^{2\alpha-4\beta}}{n^2}
\Big[
\Big(\frac{d^2\alpha}{dT^2}+\frac{d^2\beta}{dT^2}\Big)
+\Big(\frac{2d\alpha}{dT}-\frac{d\beta}{dT}-\frac{1}{n}\frac{d n}{dT}\Big)
\Big(\frac{d\alpha}{dT}+\frac{d\beta}{dT}\Big)
\Big] \nn \\
&=&-\frac{\cT F(\cT)}{4}
    \frac{\cT^3+4\tT_0 \cT^2+6\tT_0^2\cT
+2\cT+3\tT_0^3+2\tT_0-1/\tT_0}
         {\big(\cT^2+2\tT_0\cT+\tT_0^2+1\big)^4}
=-\frac{ F(\cT)\Delta(\cT)}
  {\ell^2\big(\cT^2+2\tT_0\cT+\tT_0^2+1\big)^4},
\eeq
where $\Delta$ is defined in Eq. (\ref{def_del}) and
\beq
F(\cT)
&:=&
2\cT^{6}
+12\cT^{5}{\tT_0}
+30\cT^{4}\tT_0^{2}
+5\cT^{4}
+41\cT^{3}\tT_0^{3}
+22\cT^{3}\tT_0
+\frac{\cT^{3}}{\tT_0}
+33\tT_0^{4}{\cT}^{2}
+36\,{\tT_0}^{2}{\cT}^{2}
+7{\cT}^2
\nonumber\\
&+&15{\tT_0}^{5}\cT
+24{\tT_0}^{3}\cT
+7\,{\tT_0}\cT
-2\frac{\cT}{\tT_0}
+3\tT_0^{6}
+5\tT_0^{4}
+\tT_0^{2}
-1.
\eeq
For a small $\cT$,
$$
\Omega_0^2\simeq -\frac{(1-3\tT_0^2)^2}
{4\tT_0 (1+\tT_0^2)}\cT
+\frac{(-2+25\tT_0^2+6\tT_0^3-66\tT_0^4-18\tT_0^5+27\tT_0^6)}
      {4\tT_0^2(1+\tT_0^2)^2}
+O(\cT^2),
$$
and for a large $\cT$,
\beq
\Omega_0^2=-\frac{\cT^2}{2}-\tT_0\cT +O(\cT^0).
\eeq
For a large $\cT$ $x(\cT) = -1/\cT+O(\cT^2)$, and
for a small $\cT$ the relation becomes
\begin{equation} \label{t:x 0}
x = \frac{2\big(-\tT_0\big)}{1-3\tT_0^2}
\ln \frac{\cT}{\cT_0} ,
\end{equation}
where $\cT_0(\cM/\ell)$ can be determined if we integrate out the equation~(\ref{x:t}) numerically.
Note that
$$
e^{-6\beta} = \frac{4\Delta/\ell^2}{(\cT^2+2\cT \tT_0+\tT_0^2+1)^2}
= \frac{
\cT
\big(
\cT^3+4\tT_0\cT^2+(6\tT_0^2+2)\cT
+3\tT_0^3+2\tT_0-1/\tT_0
\big)
}
{(\cT^2+2\cT \tT_0+\tT_0^2+1)^2}.
$$

\subsection{WKB solutions, matching and scalar power spectrum}

We discuss the primordial spectrum in the Bianchi IX Universe.
The new coordinate $x$ runs from $-\infty$ to $0-$, where $T\to 0+$ and $T\to
\infty$ correspond to $x=-\infty$ and $x=0-$, respectively.
The high angular-momentum limit ($J\to \infty$) corresponds
to the high momentum limit $k\gg H$ in the Bianchi I model,
since $J$ is dimensionless.

\paragraph{ Adiabatic parameter $\epsilon$ and the WKB approximation: }

We define the adiabaticity parameter,
\beq
\epsilon_{K\neq 0}:= |\epsilon|; \quad
\epsilon=
\frac{1}{|\Omega|^{3}}\frac{d\Omega^2}{dx}
=\frac{e^{\alpha-2\beta}}{n}
\frac{1}{|\Omega|^{3}}\frac{d\Omega^2}{dT}
= \frac{2\ell \Delta}{T^2+2aT+a^2+\ell^2}
\frac{1}{|\Omega|^{3}}\frac{d\Omega^2}{dT} \,.
\label{ad91}
\eeq
\begin{figure}[tbph]
\begin{center}
\begin{tabular}{ll}
\includegraphics[width=.5\linewidth,origin=tl]{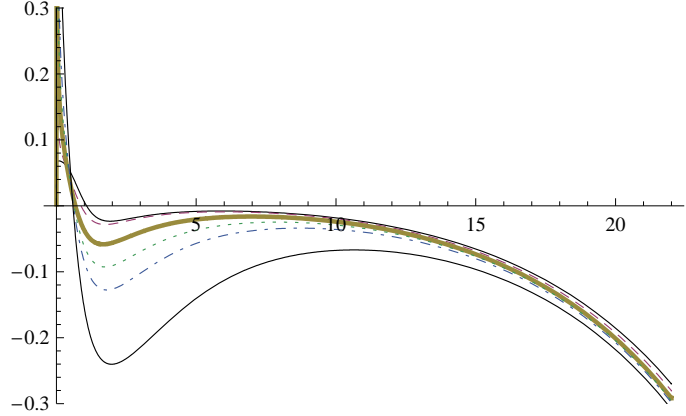}
\end{tabular}
\end{center}
\caption{The behavior of $\epsilon(\cT)$.
{The horizontal axis
represents $\cT$.}
Here we take $K=2$ and $J=30$.
The parameter $\tT_0$ 
{is}
chosen to be $\tT_0
=-0.5,-0.3,-0.1,-0.05,-0.03,-0.01$ respectively from the top.
 As seen in this figure, the large $J$ approximation gives the best result at $\cT \sim \sqrt{J}$ and gives a good approximation even for time $\cT \sim J/2$.
} \label{fig:epsilon2}
\end{figure}

We focus on the case of a small $\cT$ ($|\cT|\ll 1$).
The value of $\epsilon$ vanishes at $\cT=0$ and varies with time as
\beq \label{ep:t sim 0}
\epsilon=
\frac{\big[4\big(J(J+1)-K^2\big)+1\big]}{8\big(-\tT_0\big)^2 K^3}
\Big(1+O(\tT_0^2)\Big)\cT
-\frac{3\big[4\big(J(J+1)-K^2\big)+1\big]^2+32K^2}
      {64(-\tT_0)^3K^5}
\Big(1+O(\tT_0^2)\Big)\cT^2
+O(\cT^3).
\eeq
As in the $m=-1$ branch of the planar Kasner-de Sitter model,
the WKB approximation is always valid at the beginning
and therefore one can find a well-defined anisotropic vacuum state.
This can be understood as follows.
From Eq. (13), with use of Eqs. (6) and (12), it turns out that
in the  $T\to 0$, the initial metric functions behaves as
$n\propto T^{-1/2}$, $a\propto {\rm constant}$ and $c\propto T^{1/2}$.
After defining the proper time $d\tau=n(T)dT\approx dT/T^{1/2}$, one finds
$a\propto {\rm constant}$ and $c\propto \tau$, which
is a patch of a static spacetime as Eq. (\ref{mm}).

As $\cT$ increases $\epsilon(\cT)$ has a maximum value
$$
\epsilon_{\rm max}
\approx
\frac{\big[4\big(J(J+1)-K^2\big)+1\big]^2}
     {4(-\tT_0)K\Big[3\big[4\big(J(J+1)-K^2\big)+1\big]^2+32K^2\Big]}
\approx \frac{1}{12K(-\tT_0)},
$$
where the last step is for large $J$ at
$$
T=T_{\rm max} \simeq
\frac{4K^2\big(-\tT_0\big)\big[4\big(J(J+1)-K^2\big)+1\big]}
{3\big[4\big(J(J+1)-K^2\big)+1\big]^2+32K^2},
$$
and then start to decrease toward negative values crossing zero.
For modes satisfying $\epsilon_{\rm max} <1$,
we may use the WKB approximation {around $T=T_{\rm max}$}.
Therefore, assuming $J \gg 1$, the WKB approximation is valid at the initial
period of time for modes satisfying,
\begin{eqnarray}\label{WKB:ini}
\frac{1}{12(-\tT_0)}<K<J.
\end{eqnarray}

The subsequent behavior of the adiabatic parameter can be
described by the large $J$ approximation.
$$
\epsilon \simeq
\frac{\big(\tT_0^2+1\big)^2}
{2\tT_0 \sqrt{ J(J+1)\cT}\big(\cT^2+2\tT_0 \cT+\tT_0^2+1\big)}
\frac{3\tT_0^2+6\tT_0 \cT -1+3\cT^2}
     {\sqrt{\cT^3+4\tT_0^2\cT^2+6\tT_0^2\cT +2\cT+3\tT_0^3+2\tT_0-1/\tT_0}}.
$$
Starting from a positive value around $\cT\sim 0$, the explicit value of $\epsilon$ decreases and vanishes at
$\cT=1/\sqrt{3}-\tT_0$.
In small $\tT_0$ limit,
it continually decreases to a negative number of the order
$-1/\big(J\sqrt{-\tT_0}\big)$ and then bounces back to
zero as $\cT$ goes to infinity.
Therefore, the WKB approximation in {this}
intermediate region is valid if
\begin{equation} \label{WKB:inter}
J > \sqrt{\frac{1}{\big(-\tT_0\big)}}.
\end{equation}

This large $J$ approximation, however,
{is not satisfied}
both for small and large $\cT$ limits.
As for a small $\cT$, Eq.~(\ref{ep:t sim 0}) is a good approximation.
For a large $\cT$, the highest power terms in $\cT$ will become comparable to the large $J$ contributions, and the adiabatic parameter can be written as follows:
\begin{equation*} \label{ep:t large}
\epsilon \simeq
-\frac{\sqrt{2}
\big[1+3(-\tT_0)^{-1}(1+\tT_0^2)^2\big(J(J+1)-K^2\big)/\cT^5\big]}
{\big|2J(J+1)/(\cT^2)-1\big|^{3/2}}.
\end{equation*}
The numerator will be dominated by the $J$ dependent term for a time $\cT$ smaller than $\cT_m =(J(J+1)/\tT_0)^{1/5}$.
On the other hand, if $\cT>\cT_m$,
we may neglect the $J$ dependent term compared to {unity}.
The denominator vanishes at time $\cT_c = \sqrt{2J(J+1)}$,
leading to the divergence of the adiabaticity parameter.
For $\cT< \cT_c \sim J$,
the first term rules and for $\cT> \cT_c$,
the constant $(-1)$ rules the denominator.
Noting these {behaviors},
the order of $\epsilon$ will be minimized at time
$\cT\sim \cT_m $,
where the accuracy of the WKB approximation will be maximized.
{At a time $\cT$} satisfying $\cT_m<\cT \ll \cT_c$,
the adiabaticity parameter can be approximated to be
\begin{equation} \label{epsilon:matching}
\epsilon \simeq
-\frac{\cT^3}{2\big[J(J+1)\big]^{3/2}},
\end{equation}
which is much smaller than unity.

The adiabaticity parameter then {goes} to negative infinity at $\cT =\cT_c$
(for {a} large $J$),
and then bounces back to a finite negative {value} 
$-\sqrt{2}$ as $\cT$ goes to $\infty$.
As $\cT\to \infty$, $\epsilon$ approaches $-\sqrt{2}$
{which is larger}
than one and the WKB approximation is broken there.
In this region, we use the large $\cT$ approximation.
It is good to solve the later time evolution equation~(\ref{mode}) in $\cT$ coordinate rather than in $x$ coordinate since the order of the frequency part is transparent:
\beq \label{tOmega}
\tilde\Omega^2\equiv \frac{n^2}{e^{2\alpha}}
\Big[ \frac{J(J+1)-K^2}{e^{2\beta}} +K^2e^{4\beta}  \Big]
\approx
\frac{4J(J+1)}{\cT^4}
\Big(1-\frac{4\tT_0}{\cT}
+\frac{10\tT_0^2-2}{\cT^2}
+O(\cT^{-3})
\Big)
\,.
\eeq
The first $\cT^{-4}$ term gives the asymptotic solution $\chi_\pm$ in Eq.~(\ref{late91}) below and the second $\cT^{-5}$ term gives corrections.
Therefore, at a time $\cT \gg 1 (\gg \tT_0)$, the error of the solution $\chi_\pm$ is of order
$$
\frac{4\tT_0}{\cT} .
$$
The matching time $\cT_\ast$ is determined by the condition that this accuracy is the same as Eq.~~(\ref{epsilon:matching}):
$$
 \frac{4(-\tT_0)}{\cT}
=\frac{\cT^3}{2\big(J(J+1)\big)^{3/2}}
\,.
$$
We get the matching time
\begin{eqnarray} \label{tast:IX}
\cT_{\ast} =2\big[J(J+1)\big]^{3/8}(-\tT_0)^{1/4}.
\end{eqnarray}

In summary, the WKB approximation is valid for $\cT\ll \cT_c$ for modes satisfying Eqs.~(\ref{WKB:ini}) and (\ref{WKB:inter}).
However, it fails to be satisfied for $\cT\geq \cT_c$ and we use the
large $\cT$
approximation.
These two solutions are matched at time $\cT_\ast$ and the accuracy of the whole solution is {of $ O((-\tT_0)^{3/4}J^{-3/4})$}.

\paragraph{Moderate modes:}
If $|\tT_0|$ is smaller than $O(1)$ (${\cal M}$ is not so large),
most high momentum modes except the modes $K\leq \frac{1}{12(-\tT_0)}$
satisfy Eqs.~(\ref{WKB:ini}) and (\ref{WKB:inter}).
The main contribution {to}
the power spectrum comes from these modes.
For them, we use the WKB solution, which is given by
\beq
\label{wkb91}
\chi_{\rm WKB}(x)=\frac{1}{\sqrt{2\Omega(x)}}
    \exp \Big\{ -i \int_{x_0}^x dx' \Omega(x') +i\psi \Big\}, \quad  t\ll t_c.
\eeq

Approximating $\tilde\Omega^2$
with $4J(J+1)/\cT^4$,
the later time solution is given by
\beq
\label{late91}
&&\chi(x)\to A_+\chi_+ (x)+A_-\chi_-(x),
\nonumber\\
&&
\chi_+(x):= \frac{1}{\sqrt{2\big(J(J+1)\big)^{1/2}}}
e^{-i\big(J(J+1)\big)^{1/2}x}
\left(-1+\frac{i}{\big(J(J+1)\big)^{1/2} x}\right),\quad
\nonumber\\
&&\chi_-(x):=\frac{1}{\sqrt{2\big(J(J+1)\big)^{1/2}}}
e^{i\big(J(J+1)\big)^{1/2}x}
\left(-1-\frac{i}{\big(J(J+1)\big)^{1/2} x}\right),
\eeq
where $\chi_{+,x}\chi_--\chi_{-,x}\chi_+=-i$ and normalization condition implies the relation $|A_+|^2-|A_-|^2=1$.

The matching at $x=x_{\ast}$ determines the coefficients
\beq
A_+=-\frac{i \Phi_{\ast}}{\sqrt{2\Omega_{\ast}}}
    \Big(\chi_{-}'(x_{\ast})+i \Omega_{\ast}\chi_{-}(x_{\ast}) \Big),
\quad
A_-=\frac{i \Phi_{\ast}}{\sqrt{2\Omega_{\ast}}}
    \Big(\chi_{+}'(x_{\ast})+i \Omega_{\ast}\chi_{+}(x_{\ast}) \Big),
\eeq
where $\Phi_{\ast}$ is an unimportant phase factor in WKB solution.
The power spectrum is given by
\beq
P_{JKM}&=&\frac{1}{2\ell^2\big(J(J+1)\big)^{3/2}}
\Big|A_+-A_-\Big|^2
\nonumber\\
&=& \frac{1}{2\ell^2\Omega_\ast J(J+1)}
\Big[ \left( \sin \big[(J(J+1))^{1/2}x_\ast\big]
+\frac{ \cos \big[(J(J+1))^{1/2}x_\ast\big]}
      {(J(J+1))^{1/2}x_\ast}
- \frac{\sin \big[J(J+1)\big)^{1/2}x_\ast\big]}
       {\big[(J(J+1))^{1/2}x_\ast\big]^2}\right)^2
\nonumber\\
&+&
\frac{\Omega_\ast^2}{J(J+1)}
\left(
 \cos \big[(J(J+1))^{1/2}x_\ast\big]
-\frac{\sin \big[(J(J+1))^{1/2}x_\ast\big]}
{(J(J+1))^{1/2}x_\ast}
\right)^2\Big].
\label{psx}
\eeq
The later times the frequency squared at the present order becomes
\begin{eqnarray*}
\sqrt{J(J+1)}x_\ast \simeq -\frac{\big[J(J+1)\big]^{1/8}}
{2\big(-\tT_0\big)^{1/4}},
\quad
\Omega_\ast= \sqrt{J(J+1)}
\sqrt{1-\frac{2}{J(J+1)x_{\ast}^2} }\simeq
\sqrt{J(J+1)}.
\end{eqnarray*}
Therefore the power spectrum, including the first non-vanishing direction dependent correction, becomes
\begin{eqnarray}
P_{JKM}&\approx&
\frac{1}{2\ell^2(J(J+1))^{3/2}}
\left(
1
+\frac{8\big(-\tT_0\big)^{3/4}
\sin\big(\big[J(J+1)\big]^{1/8}(-\tT_0)^{-1/4}\big)}
      {\big[J(J+1)\big]^{3/8}}\right).
\end{eqnarray}
As the reference, it is useful to show the scalar power spectrum
in the standard de Sitter invariant vacuum is given by
\beq
P^{(0)}_{JKM}=\frac{1}{2\ell^2 (J(J+1))^{3/2}}
\eeq
and thus multiplied by the factor $\sim J^{3}$,
which is proportional to the density of state for a given $J$,
the spectrum becomes scale-invariant.
Thus, in the resultant power spectrum for the moderate modes,
the deviation from the standard spectrum
is suppressed by the factor $O(J^{-3/4})$.

\paragraph{Planar modes with $K\leq \frac{1}{12(-\tT_0)}$:}

In the case of planar modes, the WKB approximation fails to be satisfied at
$\cT\sim \cT_{\rm max}\sim
\frac{K^2(-\tT_0)}{3J^2}$.
Around these initial times, we try to find an approximate solution for the times $t \ll \tT_0$.
We need to 
expand the frequency squared up to first order in $\tT_0 \cT$:
\begin{eqnarray}\label{tilde Om:0}
{\tilde\Omega}^2 &\simeq&
\frac{4\tT_0^2}
     {\big(1-3\tT_0^2\big)^2 \cT^2}
\left(K^2
+ \frac{J(J+1)(1-3\tT_0^2)\cT}
       {\big(1+\tT_0^2\big)\big(-\tT_0\big)}
\right) ,
\end{eqnarray}
where we ignore $1/\tT_0$ and $K$ relative to $J$.
Noting $\frac{d(3\alpha-\log n)}{dT}\simeq \frac{1}{T}$,
we get the solution to the differential equation~(\ref{mode}),
\begin{eqnarray} \label{sol1:planar}
\phi_1(\cT) =
\sqrt{\frac{2\big(-\tT_0\big)}
{\big(1-3\tT_0^2\big)
   \sinh\big(\pi q_K\big)}}\,
 J_{-i q_K}(A_J \cT^{1/2}) ,
\end{eqnarray}
where
\beq
q_K:=\frac{4(-\tT_0)}{1-3\tT_0^2}K,\quad
A_J=4\sqrt{\frac{J(J+1)(-\tT_0)}
      {(1-3\tT_0^2)\big(1+\tT_0^2\big)}},
\eeq
and we choose the normalization of the solution so that it becomes an incoming wave 
in $x$ coordinate at $x \to -\infty$,
noting the relation~(\ref{t:x 0}) between $\cT$ and $x$.
If we use the solution~(\ref{sol1:planar}) up to a time $\cT_1$,
the accuracy of the above solution will be of order
$\cT_1 (-\tT_0)$.

On the other hand, the WKB solution is still valid at a time $\cT_1$ satisfying $ \frac{2\big(-\tT_0\big)^2K^2}{J^2}\ll \cT_1 \ll 1$.
At this time we may also use the high $J$ limit and the adiabaticity parameter becomes $\epsilon \simeq \frac{1}{2(-\tT_0)^{1/2}J\cT^{1/2}}$.
The matching time $\cT_1$ between the initial exact solution and the WKB solution can be determined by setting the accuracies to be the same:
$$
\cT_1(-\tT_0) = \frac{1}{2\sqrt{-\tT_0} J \cT_1^{1/2}} .
$$
Therefore, the matching time becomes
$\displaystyle \cT_1 = \frac{1}{2^{2/3}(-\tT_0)J^{2/3}}$.
The WKB solution becomes
\begin{eqnarray}
\phi_2(\cT) = e^{-\alpha-\beta} \chi_{\rm WKB}(x)
=\sqrt{\frac{2}{\Omega\big(T^2+2t_0 T+t_0^2+\ell^2\big)}} \left( B_+
    e^{-i \int_{T_1}^T \Omega(T') \frac{dx(T')}{dT'} dT'} +  B_-
    e^{i \int_{T_1}^T \Omega(T') \frac{dx(T')}{dT'} dT'} \right).
\end{eqnarray}
We use this solution during the period of time
$\cT_1<\cT<\cT_\ast$.
The coefficient $B_\pm$ is determined to be
\begin{eqnarray}
B_\pm &=&
\frac{\sqrt{\Omega_1\big(1+\tT_0^2\big)}}
     {2\sqrt{2}}
 \left(\phi_1(T_1) \pm
\frac{i(1-3\tT_0^2) T_1} {2(-\tT_0) \Omega_1}
\frac{d\phi_1(t_1)}{dT} \right)
=\frac{1}{2}\sqrt{\frac{(1+\tT_0^2)e^{-\pi q_K(1\mp 1)}}
 {1- e^{-2\pi q_K}}}
 \cdot e^{\mp i(A_{J}\cT_1^{1/2}
         -\frac{\pi}4)},
\end{eqnarray}
where
we write down the zeroth order term only.

After {$\cT>\cT_\ast$}, we may once again can use the asymptotic solution $\chi_\pm$ to compare it with the WKB solution $\chi_{\rm WKB}$.
{The late time solution is given by}
$$
\chi(x) =  c_+\chi_+(x) +  c_-\chi_-(x).
$$
At {$\cT_\ast$}, $\Omega_\ast = J(J+1)$ and
$(J(J+1))^{1/2}x_\ast \simeq -J^{1/4}/(-\tT_0)^{1/4}$.

The primordial power spectrum is proportional to the square of the size of $c_+-c_-$.
{Matching at the boundary $x_\ast$}, we have
$$
c_+ = - i \left(\chi_{\rm WKB} \frac{d\chi_-}{dx}- \frac{d\chi_{WKB}}{dx} \chi_-\right)_{x=x_\ast},\quad
c_- =  i \left(\chi_{\rm WKB} \frac{d\chi_+}{dx}- \frac{d\chi_{WKB}}{dx} \chi_+\right)_{x=x_\ast}.
$$
If we keep {only the zeroth order}, the power spectrum becomes
\beq
P_{JKM}=\frac{1}{2\ell^2\big(J(J+1)\big)^{3/2}}|c_+-c_-|^2
= \frac{1+\tT_0^2}{2 \ell^2 \big(J(J+1)\big)^{3/2}}
\left(\frac{1+ e^{-2\pi q_K}}{1-e^{-2\pi q_K}}
- \frac{2e^{-\pi q_K}}{1-e^{-2\pi q_K}}
   \cos 2\Phi
\right),
\eeq
where $\Phi = (J(J+1))^{1/2}x_\ast
+ \int_{x_1}^{x_\ast} \Omega dx
+A_J\cT_1^{1/2}-\frac{\pi}4$ and
$q_K=4(-\tT_0)K/(1-3\tT_0^2)$
contains the direction dependence.
Thus, the explicit direction dependence is not suppressed.
Note that $\Psi\sim J$ rules the phase factor $\Phi$ for larger $J$.
The $K=0$ limit is not well defined,
however, {this mode is not in
our concern} since the wavelength of
such a perturbation mode is beyond our Hubble
horizon.

\section{Conclusion and discussions}

In this article, we considered
the quantization of a massless and minimally coupled scalar field
in the Universe,
which is initially anisotropic
and approaches the de Sitter {spacetime}.
The motivation to consider the initially anisotropic Universe
is two-fold:
The first motivation is that even if the current Universe is almost isotropic, it does not mean that the Universe is isotropic from the beginning.
In fact, Wald's no-hair theorem ensures that in the presence of a positive cosmological constant
an initially anisotropic Universe exponentially approaches a
de Sitter spacetime at the later time
under the strong or dominant energy condition.
Therefore,
it would be more generic that the initial Universe is anisotropic.
The second motivation comes from the recent observations by WMAP satellite.
WMAP measured the temperature fluctuations of cosmic microwave background (CMB)
and almost confirmed the predictions from the inflation, during which the Gaussian and statistically isotropic
primordial fluctuations are produced.
But after the WMAP data were released, several groups have reported the so-called low-$\ell$ anomalies
in large angular power of CMB fluctuations, e.g.,
the suppression of the power of quadrupole,
the planarity of quadrupole and octopole CMB maps,
the alignment of the preferred directions of quadrupole and octopole moments,
and so on.
They may not be satisfactorily explained by the standard isotropic initial state,
but may be done by the direction-dependent primordial fluctuations.

In this paper,
we considered the gravitational theory
composed of the Einstein-Hilbert term and a positive cosmological constant.
For simplicity, we assume that the late-time inflationary stage
is exactly described by the de Sitter solution
and ignored the dynamics of an inflaton field for simplicity.
We considered two kinds of initially anisotropic Universe, say,
the Bianchi I and Bianchi IX models.
In each model, there is {an exact solution}.
In the Bianchi I model,
we considered the Kasner-de Sitter solution,
in which the spacetime geometry
is initially a Kasner spacetime and approaches the de Sitter spacetime.
In the Bianchi IX model, similarly we considered
the Taub-NUT de Sitter solution,
in which the spacetime geometry is initially a Taub-NUT spacetime.
Note that in the case of Bianchi I,
we focused on the case that the Universe has an exact planar symmetry,
which is isotropic along two of three spatial axes.
In the Bianchi IX model,
the exact solution has only two independent scale factors.

After giving the background geometry, we investigated the spectrum of a scalar field,
which is the counterpart of the inflaton fluctuation.
We discussed how a massless scalar field is quantized in the initial anisotropic stage.
The qualitative behaviors of the resultant spectrum in both the Bianchi I and IX models
are very similar and therefore, here we summarize our result mainly focusing on the Bianchi I model.
We found that in the case with a planar symmetry,
there is a well-defined adiabatic vacuum unless $k_3\neq 0$,
where $k_3$ is the comoving momentum along the preferred direction.
As we mentioned previously,
in a quantum harmonic oscillator system,
an {\it adiabatic} process usually implies the
one where the potential changes slowly enough compared to its size,
and the time evolution can be obtained from the zero-th order
WKB approximation.
We followed this definition for the term {\it adiabatic}. 
In the standard inflationary models,
an adiabatic vacuum is also defined in the same way. 
In our case
an adiabatic vacuum state, called an anisotropic vacuum in this paper,
was found only
in the special solutions of anisotropic Universe,
which are regular in the initial times.
It was shown that for the moderate modes, $k_3\sim k$,
where $k$ is the total comoving momentum, the scalar power spectrum has an oscillatory behavior in the
smaller value of $k$ and a {suppression}
of a large scale power.
For the planar mode, $k_3 \ll  k$, the adiabaticity parameter vanishes in the earlier times
the scalar power spectrum is well defined,
but during the intermediate times
it becomes greater than unity.
Then,
the effect
of primordial anisotropy is enhanced and in the resultant
spectrum
the scale-dependence is unsuppressed.
For the modes of $k_3=0$, the adiabaticity parameter diverges and
the WKB approximation is not well defined.
But such a mode is not observable.

In the case of the Bianchi IX model, by definition, the modes are discrete, but
each mode exhibits a similar behavior to the corresponding mode in the case
of Bianchi I model.
For the moderate modes that satisfy $K>\ell/(-12t_0)$,
where the quantum number $K$ characterizes the angular momentum along
the preferred direction ($\sim k_3/H$ in the Bianchi I model),
in the resultant power spectrum the angular dependence is suppressed.
Note that roughly speaking $t_0$ characterizes the degree
of the initial anisotropy,
and $\ell$ is related to the late time expansion rate of de Sitter
($\ell=1/H$).
In contrast,
for the planar modes $K<\ell/(-12t_0)$,
the effect of the primordial anisotropy is not suppressed.

In summary,
the anisotropy is not always suppressed in the present models.
The anisotropy of the planar modes may leave non-negligible
effects on the plane orthogonal
to the preferred direction.
They would be generic predictions from an initially anisotropic
Universe in vacuum.
There are important issues left for future studies.
To obtain more reliable predictions,
our method should be applied to
solve the metric perturbations
and in particular
to find out the role of coupling of one of the tensor polarizations
to the scalar mode.
The evaluation of the bispectrum or trispectrum would also be
important.

\section*{Acknowledgement}
This work was supported by the Korea Research Foundation Grant funded by the Korean government (MOEHRD) KRF-2008-314-C00063 (HCK) and by the National Research Foundation of Korea (NRF) grant funded by the Korean government (MEST)( No. 20090063070, MM).

\appendix

\section{Evolution of Bianchi IX Universe}

Employing the Hamiltonian formalism, one can study the
cosmology in the Bianchi universe in analogy with
the classical mechanics.
By parameterizing the scale factors of three independent
spatial directions as
$e^{\alpha+\beta_++\sqrt{3}\beta_-}$,
$e^{\alpha+\beta_+-\sqrt{3}\beta_-}$
and $e^{\alpha-2\beta_+}$,
the comoving evolution
can be specified by two variables $(\beta_+,\beta_-)$.
In analogy with the classical mechanism, the evolution of
the Bianchi IX Universe can be described by
the motion of a particle on the $(\beta_+,\beta_-)$ plane
with the potential
\beq
V_{\rm IX}(\beta_+,\beta_-)
=2e^{4\beta_+}\cosh\big(4\sqrt{3}\beta_-\big)
+e^{-8\beta_+}
-2e^{4\beta_+}
-4e^{-2\beta_+}\cosh\big(2\sqrt{3}\beta_-\big).
\label{IX}
\eeq
The potential is almost flat near the origin $(\beta_+=0,\beta_-=0)$
but there is a steep exponential wall which forms an equilateral.
Note that
the corresponding potential in the Bianchi I model is vanishing,
say $V_{\rm I}(\beta_+,\beta_-)=0$.
The cosmic motion is almost confined inside this wall.
There are three valleys into the potential
along the straight lines,
$\beta_-=0$ ($\beta_+>0$),
$\beta_-=\sqrt{3}\beta_+$ and $\beta_{-}=-\sqrt{3}\beta_+$ (for
$\beta_+<0$).
Only along these valleys, the Bianchi IX Universe can evolve to have highly anisotropic geometry with
large $\beta_+$ or $\beta_-$.
These three valleys are equivalent under the exchange of label of the axes.

The evolution {of} Taub-NUT de Sitter spacetime is initially along one of these valleys
on the $(\beta_+,\beta_-)$ plane, starting from the infinity.
In fact, one can read off the correspondence of the variables
$\alpha + \beta_+ + \sqrt{3}\beta_-=\ln a$,
$\alpha + \beta_+ - \sqrt{3}\beta_-=\ln a$,
and thus
$\alpha -2\beta_+=\ln c,$
and one can easily find
$\alpha=(1/3)\ln (a^2c)$,
$\beta_+=(1/3)\ln\big(a/c\big)$
and
$\beta_-=0$.

\end{document}